\documentclass[aps,pra,floatfix,twocolumn,superscriptaddress]{revtex4-1}
\usepackage{amsmath,amssymb}
\usepackage{graphicx}
\usepackage{epsfig}
\usepackage{psfrag}
\usepackage[usenames]{color}
\usepackage{xcolor}
\usepackage{hyperref}
\usepackage{soul,color}
\usepackage{tabu}
\usepackage{multirow}
\hypersetup{colorlinks=true,citecolor={blue},linkcolor={blue},urlcolor={blue}}

\begin{document}

\title{Theory of BCS-like bogolon-mediated  superconductivity \\in transition metal dichalcogenides}

\author{Meng~Sun}
\affiliation{Center for Theoretical Physics of Complex Systems, Institute for Basic Science (IBS), Daejeon 34126, Korea}
\affiliation{Basic Science Program, Korea University of Science and Technology (UST), Daejeon 34113, Korea}

\author{A.~V.~Parafilo}
\affiliation{Center for Theoretical Physics of Complex Systems, Institute for Basic Science (IBS), Daejeon 34126, Korea}

\author{K.~H.~A.~Villegas}
\affiliation{Center for Theoretical Physics of Complex Systems, Institute for Basic Science (IBS), Daejeon 34126, Korea}
\affiliation{Division of Physics and Applied Physics, Nanyang Technological University, 637371 Singapore, Singapore}

\author{V.~M.~Kovalev}
\affiliation{A.~V.~Rzhanov Institute of Semiconductor Physics, Siberian Branch of Russian Academy of Sciences, Novosibirsk 630090, Russia}
\affiliation{Novosibirsk State Technical University, Novosibirsk 630073, Russia}

\author{I.~G.~Savenko}\email[Corresponding author: ]{ivan.g.savenko@gmail.com}
\affiliation{Center for Theoretical Physics of Complex Systems, Institute for Basic Science (IBS), Daejeon 34126, Korea}
\affiliation{Basic Science Program, Korea University of Science and Technology (UST), Daejeon 34113, Korea}

\date{\today}

\begin{abstract}
We report on a novel mechanism of BCS-like superconductivity, mediated by a pair of Bogoliubov quasiparticles (bogolons). 
It takes place in hybrid systems consisting of a two-dimensional electron gas in a transition metal dichalcogenide monolayer in the vicinity of a Bose-Einstein condensate.
Taking a system of two-dimensional indirect excitons as a testing ground we show, that the bogolon-pair-mediated electron pairing mechanism is stronger than phonon-mediated and single bogolon-mediated ones.
We develop a microscopic theory of bogolon-pair-mediated superconductivity, based on the Schrieffer--Wolff transformation and the Gor'kov's equations, study the temperature dependence of the superconducting gap and estimate the critical temperature of superconducting transition for various concentrations of the electron gas and the condensate densities. 
\end{abstract}	

\maketitle


\section{Introduction}
The conventional microscopic Bardeen-Cooper-Schrieffer (BCS) superconductivity originates from the interaction between electrons and phonons (crystal lattice vibrations), which results in the attraction between electrons with opposite momenta and spins with the sequential formation of Cooper pairs~\cite{RefMainBCS, PhysRev.108.1175}. 
However, this phenomenon is usually observed at low temperatures (as compared with room temperature), of the order of several Kelvin since the phonon-mediated superconducting (SC) gap usually amounts to several meV. 
And superconductors with the critical temperature of SC transition $T_c$ above 30~K are traditionally considered high-temperature superconductors~\cite{Bednorz1986}.

In an attempt to increase the electron-phonon coupling and $T_c$, one immediately faces certain  obstacles, one of which is the Peierls instability~\cite{RevModPhys.60.1129}.
In the mean time, the search for high-temperature superconductivity is a rapidly developing area of research nowadays, especially in low-dimensional systems~\cite{NatureBilayer, PhysRevLett.122.027001}. 
In hybrid superconductor-semiconductor electronics and circuit quantum electrodynamics, 
two-dimensional (2D) superconductors might allow for scaling down the characteristic size of a device down to atomic-scale thickness for possible application in quantum computing~\cite{Frasca2019, Burkard2020, PhysRevLett.124.087701}. 
Low-dimensional superconductors also provide such advantages as the robustness against in-plane magnetic fields due to the spin-valley locking~\cite{Saito2016} and an additional enlargement of $T_c$ in the atomic-scale layer limit~\cite{Ge2015}. 
From the fundamental side, the SC phase in samples of lower dimensionality usually either co-exists or competes with other (coherent) many-body phases such as the quantum metallic or insulator states, the charge density wave, or magnetic phase, giving rise to richer physics than in three-dimensional systems~\cite{Uchihashi2016}. 
The drawbacks and limitations of phonons as mediators of electron pairing for realizing high-$T_c$ 2D superconductors motivate the search for other pairing mechanisms.

There have been various attempts to replace regular phonons by some other quasiparticles aiming at increasing $T_c$ and the SC gap. 
One of the routes is exciton-mediated superconductivity~\cite{little, ginzburg, PhysRevB.7.1020}.
Photon-mediated superconductivity has also been recently predicted~\cite{PhysRevLett.122.133602}.
Another way is to use the excitations above a Bose-Einstein condensate (BEC), called the Bogoliubov quasiparticles (bogolons) in hybrid Bose-Fermi systems, where one expects the SC transition in the fermionic subsystem. 
The bosonic subsystem can be represented by an exciton or exciton-polariton condensate, which have been predicted~\cite{PhysRevA.53.4250, lozovik, Fogler2014, PhysRevB.92.165121, PhysRevB.93.245410, PhysRevB.96.174504} and studied experimentally~\cite{Room1, Room2, WangNature2019} at relatively high temperatures sometimes reaching the room temperature. 
In systems of indirect excitons, spatially separated electron-hole pairs, achieving high-temperature condensation should be possible if using 2D materials based on transition metal dichalcogenides such as MoS$_2$ thank to large exciton binding energy~\cite{doi:10.1021/nl501133c}.
Bogolons possess some of the properties of acoustic phonons and can, in principle, give electron pairing, as it has been theoretically shown in several works~\cite{Laussy:2010aa, Cotleifmmode-telse-tfi:2016aa, Skopelitis:2018aa}.
These proposals, however, operated with single-particle (single-bogolon) pairing, assuming that multi-particle processes belong to the higher orders of the perturbation theory and thus they are weak and can be safely disregarded. 
Is this widespread assumption true?

As the earlier work~\cite{Villegas2019} points out, the bogolon-pair-mediated processes (\textit{2b} processes in what follows) can give the main contribution when considering the scattering of electron gas in the normal state (above $T_c$). 
If we go down $T_c$, several questions arise naturally.
Will there occur 2b-mediated pairing? 
What is its magnitude, as compared with single-bogolon (\textit{1b}) processes? 
Is the parameter range (in particular, condensate density, concentration of electrons in 2DEG) achievable experimentally?
In this article, using the BCS formalism we develop a microscopic theory of {2b} superconductivity and address all these questions.


\section{Theoretical framework}
\begin{figure}[!t]
\includegraphics[width=0.49\textwidth]{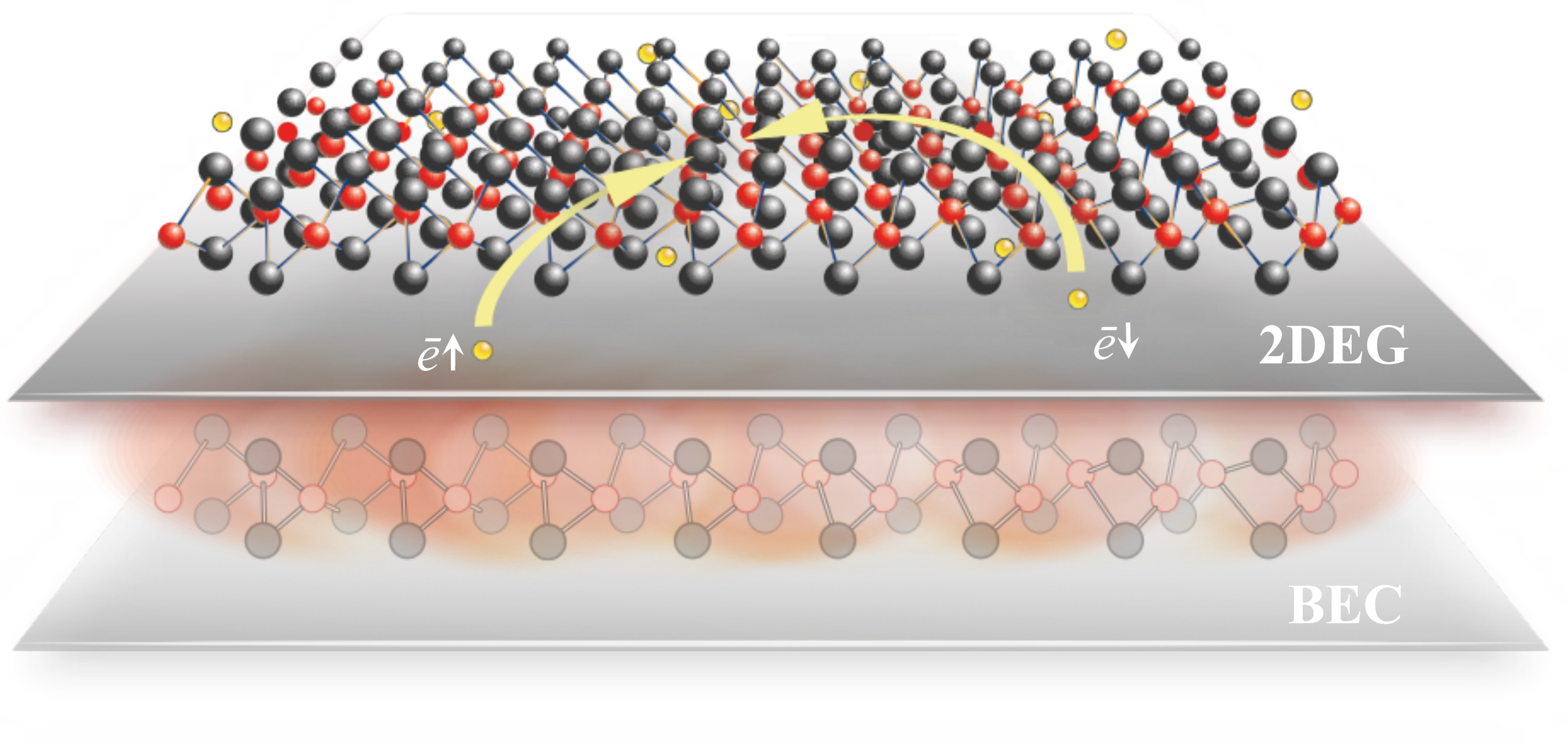}
\caption{System schematic.
Two-dimensional electron gas (2DEG) with parabolic dispersion in the vicinity of a 2D Bose-Einstein condensate (BEC).
We consider the BEC of indirect excitons, which reside in a double quantum well: n-doped and p-doped layers of MoS$_2$ and WSe$_2$  separated by an hBN.
Electrons and the condensate particles are coupled via the Coulomb forces, which allows electrons with opposite spins (yellow dots) form Cooper pairs.
}
\label{Fig1}
\end{figure}

Let us consider a hybrid system consisting of a 2D electron gas (2DEG) and a 2D Bose-Einstein condensate (BEC), taking indirect excitons as an example, where the formation of BEC has been reported~\cite{Butov2017, WangNature2019} (Fig.~\ref{Fig1}).
The electrons and holes reside in n- and p-doped layers, respectively.
These layers can be made of MoS$_2$ and WSe$_2$ materials separated by several layers of hexagonal boron nitride (hBN)~\cite{WangNature2019}.
The 2DEG and exciton layers are also spatially separated by hBN and the particles are coupled by the Coulomb interaction~\cite{Boev:2016aa, Matuszewski:2012aa} described by the Hamiltonian
\begin{equation}\label{eq.1}
{\cal H}=\int d\mathbf{r}\int d\mathbf{R}\Psi^\dag_\mathbf{r}\Psi_\mathbf{r}g\left(\mathbf{r}-\mathbf{R}\right)\Phi^\dag_\mathbf{R}\Phi_\mathbf{R},
\end{equation}
where $\Psi_\mathbf{r}$ and $\Phi_\mathbf{R}$ are the field operators of electrons and excitons, respectively, $g\left(\mathbf{r}-\mathbf{R}\right)$ is the strength of Coulomb interaction between the particles, $\mathbf{r}$ and $\mathbf{R}$ are the in-plane coordinates of the electron and the exciton center-of-mass motion.

Furthermore, we assume the excitons to be in the BEC phase. Then we can use the model of a weakly interacting Bose gas and  split $\Phi_\mathbf{R}=\sqrt{n_c}+\varphi_\mathbf{R}$, where $n_c$ is the condensate density and $\varphi_\mathbf{R}$ is the field operator of the excitations above the BEC.
Then, the Hamiltonian~\eqref{eq.1} breaks into three terms, two of which are
\begin{eqnarray}
\label{EqTwoHam11}
{\cal H}_{1}&=&\sqrt{n_c}\int d\mathbf{r}\Psi^\dag_\mathbf{r}\Psi_\mathbf{r} \int d\mathbf{R}g\left(\mathbf{r}-\mathbf{R}\right)\left[\varphi^\dag_\mathbf{R}+\varphi_\mathbf{R}\right],\\
\label{EqTwoHam12}
{\cal H}_{2}&=&\int d\mathbf{r}\Psi^\dag_\mathbf{r}\Psi_\mathbf{r}\int d\mathbf{R}g(\mathbf{r}-\mathbf{R})\varphi^\dag_\mathbf{R}\varphi_\mathbf{R}.
\end{eqnarray}
The first term, ${\cal H}_{1}$, is responsible for electron-single bogolon interaction, and the second term, ${\cal H}_{2}$, is bogolon-pair-mediated.
The third term reads $gn_c\int d\mathbf{r}\Psi^\dag_\mathbf{r}\Psi_\mathbf{r}$. 
It gives a shift $\delta\mu= gn_c$ of the Fermi energy  $\mu=\hbar^2p_F^2/2m$, where $p_F$ is the Fermi wave vector and $m$ is electron effective mass. 
Then $p_F$ also becomes $n_c$-dependent, strictly speaking, but we disregard this correction in what follows. 

We express the field operators as the Fourier series,
\begin{eqnarray}
\nonumber
\varphi_\mathbf{R}
=\frac{1}{{L}}
\sum_{\mathbf{p}} e^{i\mathbf{p}\cdot\mathbf{R}}
(u_\mathbf{p}b_\mathbf{p}+v_{\mathbf{p}}b^\dagger_\mathbf{-p}),~
\Psi_\mathbf{r}&=&\frac{1}{{L}}\sum_\mathbf{k}e^{i\mathbf{k}\cdot\mathbf{r}}c_\mathbf{k},~
\end{eqnarray}
where $b_{\mathbf{p}}$($c_\mathbf{k}$) and $b^\dag_{\mathbf{p}}$($c^\dagger_\mathbf{k}$) are the bogolon (electron) annihilation and creation operators, respectively, and $L$ is the length of the sample. 
The Bogoliubov coefficients read~\cite{Giorgini:1998aa}
\begin{eqnarray}\label{eq.4}
&&u^2_{\mathbf{p}}=1+v^2_{\mathbf{p}}=\frac{1}{2}\left(1+\left[1+\left(\frac{Ms^2}{\omega_{\mathbf{p}}}\right)^2\right]^{1/2}\right),
\\
\nonumber
&&~~~~~u_{\mathbf{p}}v_{\mathbf{p}}=-\frac{Ms^2}{2\omega_{\mathbf{p}}},
\end{eqnarray}
where $M$ is the exciton mass, $s=\sqrt{\kappa n_c/M}$ is the sound velocity, $\kappa=e_0^2d/\epsilon_0\epsilon$ is the exciton-exciton interaction strength in the reciprocal space, $e_0$ is electron charge, $\epsilon$ is the dielectric constant, $\epsilon_0$ is the dielectric permittivity,
$\omega_p=\hbar sp(1+p^2\xi_h^2)^{1/2}$ is the spectrum of bogolons, and $\xi_h=\hbar/2Ms$ is the healing length. %
Then Eqs.~\eqref{EqTwoHam11} and~\eqref{EqTwoHam12} transform into
\begin{eqnarray}
\label{Eq1bexpr}
{\cal H}_1&=&
\frac{\sqrt{n_c}}{L}
\sum_{\bf k, p}
g_{\bf p}
\left[(v_{\bf p}+u_{\bf -p})b^{\dag}_{-\bf p}
\right.
\\
\nonumber
&&\left.~~~~~~~~~~~~~~~~~~~~+(v_{\bf -p}+u_{\bf p})b_{\bf p}\right]c^{\dag}_{\bf k+p}c_{\bf k},\\
\label{Eq2b}
{\cal H}_2&=&
\frac{1}{L^2}
\sum_{\bf k, p, q}
g_{\bf p}
\left[
u_{\bf q-p}u_{\bf q}b^{\dag}_{\bf q-p}b_{\bf q}+u_{\bf q-p}v_{\bf q}b^{\dag}_{\bf q-p}b^{\dag}_{\bf -q}\right.\\
\nonumber
&&\left.+v_{\bf q-p}u_{\bf q}b_{\bf -q+p}b_{\bf q}
+v_{\bf q-p}v_{\bf q}b_{\bf -q+p}b^{\dag}_{\bf -q}
\right]
c^{\dag}_{\bf k+p}c_{\bf k},
\end{eqnarray}
where $g_p$ 
is the Fourier image of the electron-exciton interaction. 
Disregarding the peculiarities of the exciton internal motion (relative motion of the electron and hole in the exciton), we write the electron-exciton interaction in direct space as
\begin{eqnarray}
\label{EqgDirSp}
g(\mathbf{r} - \mathbf{R} ) &=& \frac{e_0^2}{4\pi \epsilon_0 \epsilon} \left( \frac{1}{r_{e-e}} - \frac{1}{r_{e-h}} \right),
\end{eqnarray}
where $r_{e-e}=\sqrt{l^2 +(\mathbf{r}-\mathbf{R})^2}$  and $r_{e-h}= \sqrt{(l+d)^2 + (\mathbf{r}-\mathbf{R})^2}$;
$d$ is an effective size of the boson, which is equal to the distance between the n- and p-doped layers in the case of indirect exciton condensate, 
and $l$ is the separation between the 2DEG and the BEC~\cite{[{If we assume the contact interaction, taking small $d$ and $l$ thus $pd,~pl\ll1$ for all $p$, then we can use $g_p\approx e^2_0d/2\epsilon_0\epsilon$ treating electron-boson interaction in exciton or exciton-polariton condensates~\cite{Laussy:2010aa}.
However, this is not always the case, and in what follows, we use the general interaction term}]C1}.
The Fourier transform of~\eqref{EqgDirSp} gives
\begin{equation}
    g_p=\frac{e^2_0 \left(1-e^{-pd}\right)e^{-pl}}{2\epsilon_0\epsilon p}.
\end{equation}

Following the BCS approach~\cite{Mahan}, we find the effective electron  s-wave~\cite{[{Despite the fact that we do not explicitly write the spin indeces in Eqs.~(\ref{Eq1bexpr})-(\ref{EqPairingHam2b}), we investigate the case of s-wave pairing, when two electrons in the Cooper pair have opposite momenta and spins. In other words, the Cooper pair is in a spin-singlet state with $S=0$ and $L=0$, where $S$ and $L$ are the spin and orbital momenta, respectively. This assumption is legitimate since the Fermi surface of the 2DEG and bogolon-mediated interaction are isotropic}]C4} pairing Hamiltonian (see Supplemental Material~\cite{[{See Supplemental Material at [URL], which gives the details of the derivations of main formulas}]SMBG}),
considering 1b and 2b processes separately to simplify the derivations and draw the comparison between them,
\begin{eqnarray}
\label{EqPairingHam2b}
{\cal H}^{(\lambda)}_\textrm{eff}={\cal H}_0+\frac{1}{2L^2}
\sum_{\bf k,k',p}
V_{\lambda}(p)
c^{\dag}_{\bf k+p}c_{\bf k}c^{\dag}_{\bf k'-p}c_{\bf k'},
\end{eqnarray}
where ${\cal H}_0$ is a free particle dispersion term and
\begin{eqnarray}
\label{EqEffPot1b}
V_{1b}(p)&=&-\frac{n_c}{Ms^2}g_p^2,
\\
\label{EqEffPot2b}
V_{2b}(p)&=&-
\frac{M^2s}{4\hbar^3}
\frac{g_p^2}{p}
\left(
1+
\frac{8}{\pi}
\int\limits_{p_\textrm{min}}^{p/2 } 
\frac{dqN_q}{\sqrt{p^2-4q^2}}
\right)
\end{eqnarray}
are effective potentials of electron-electron interaction.
In Eq.~\eqref{EqEffPot2b}, $N_q=\left[\exp(\frac{\omega_q}{k_BT})-1\right]^{-1}$ is the bogolon Bose distribution function. 
It gives the divergence of the integral at $q=0$ typical for 2D systems~\cite{Hohenberg1967, Bagnato1991, Butov2017}. 
Therefore, we introduce a cutoff $p_\textrm{min}$, responsible for the convergence and associated with the finite size of the sample (or condensate trapping).
The factor $N_q$ emerges at finite temperatures and gives an increase of the exchange interaction between electrons. 
The number of thermally activated bogolons increases with temperature, which enhances the 2b-mediated electron scattering.

Furthermore, we use 
the equation for the SC gap $\Delta_\lambda$~\cite{Mahan}
\begin{eqnarray}\label{selfconsistent}
\Delta_\lambda(\mathbf{k})=-\frac{1}{L^2}
\sum_\mathbf{p}V_\lambda(p)\frac{\Delta_\lambda(\mathbf{k-p})}{2\zeta^{(\lambda)}_{\mathbf{k}-\mathbf{p}}}\tanh\left(\frac{\zeta^{(\lambda)}_{\mathbf{k}-\mathbf{p}}}{2k_BT}\right),~~~
\end{eqnarray}
where $\zeta^{(\lambda)}_\mathbf{k}=\sqrt{\xi^2_\mathbf{k}+\Delta_\lambda^2(\mathbf{k})}$ with $\xi_\mathbf{k}=\hbar^2k^2/2m-\mu$ being the kinetic energy of particles measured with respect to the Fermi energy. Then, we change the integration variable and cancel out $\Delta_{\lambda}$ in both sides of Eq.~(\ref{selfconsistent}) [since we consider the s-wave pairing when the SC gap is momentum independent]. 
As a result, Eq.~(\ref{selfconsistent}) transforms into
\begin{eqnarray}\label{EqGap11}
1=-
\int_0^{\infty} \frac{dp p}{2\pi}\int_0^{2\pi}\frac{d\theta}{2\pi} \frac{V_\lambda(|\mathbf{k-p}|)}{2\zeta^{(\lambda)}_{\mathbf{p}}}\tanh\left(\frac{\zeta^{(\lambda)}_{\mathbf{p}}}{2k_BT}\right),~~~
\end{eqnarray}
where $\theta$ is the angle between the vectors $\mathbf{k}$ and $\mathbf{p}$.
Furthermore, we switch from the integration over the momentum to the integration over the energy: $p\rightarrow 2m(\mu+\xi)$, and introduce an effective cut-off $\omega_b=\hbar s/\xi_h$ in accordance with the BCS theory. 
This parameter appears by analogy with the Debye energy $\omega_D$ (in the case of acoustic phonon-mediated pairing), which is connected with the minimal sound wavelength of the order of the lattice constant and has obvious physical meaning.
In the case of bogolons, this cut-off is less intuitive and, in principle, it remains a phenomenological parameter~\cite{Skopelitis:2018aa}. 
Its value $\hbar s/\xi_h$ might be attributed to the absence of bogolon excitations with wavelengthes shorter than the condensate healing length.



Let us, first, consider zero-temperature case, when the $\tanh$ function in Eq.~(\ref{EqGap11}) becomes unity and $N_q=0$. 
Assuming that the main contribution into the effective electron-electron interaction comes from electrons near the Fermi surface and $p_F d, p_F l\ll 1$, we find analytical expressions,
\begin{eqnarray}
\label{Gap1bog}
\Delta_{1b}(T=0)=2\omega_b \exp\left[-\frac{8 Ms^2}{\nu_0  n_c }\left(\frac{\epsilon_0\epsilon }{e_0^2d}\right)^2\right],\\
\label{Gap2bog}
\Delta_{2b}(T=0)=2\omega_b \exp\left[-\frac{16 \hbar^3 p_F}{\tilde{\nu}_0 M^2s}\left(\frac{\epsilon_0\epsilon }{e_0^2d}\right)^2\right],
\end{eqnarray}
where $\nu_0=m/\pi \hbar^2$ is a density of states of 2DEG, $\tilde{\nu}_0=\nu_0 \log(4p_F L)/\pi$ is an effective density of states, and $L$ is the system size. 
Note, that in Eq.~(\ref{Gap2bog}) there emerges an additional logarithmic factor (as compared with the standard BCS theory).
It happens due to the momentum dependence of the 2b-mediated pairing potential $V_{2b}$ and due to the integration over the angle $\theta$ in the self-consistent equation for the SC gap [Eq.~(\ref{EqGap11})].

\begin{figure}[!t]
    \centering
    \includegraphics[width=0.47\textwidth]{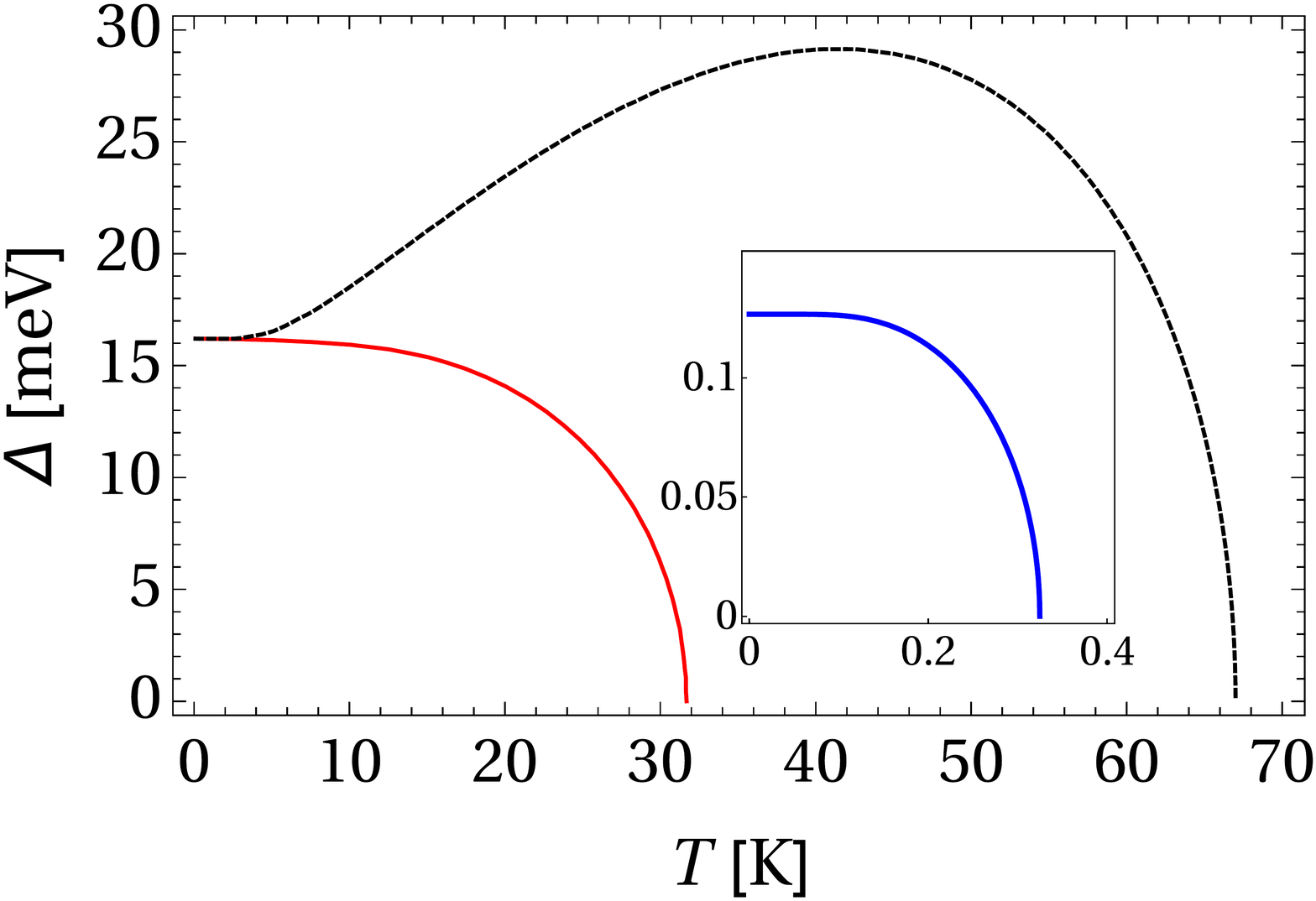}
    \caption{Superconducting gap as a function of temperature.
    Red solid curve shows 2b-mediated gap disregarding $N_q$-containing term in Eq.~\eqref{EqEffPot2b}.
    Black dashed curve accounts for the full temperature dependece (including the influence of $N_q$-containing term in Eq.~\eqref{EqEffPot2b}).
    Inset shows one-bogolon SC gap for comparison.
    We used the parameters, typical for MoS$_2$ and hBN:
    $\epsilon=4.89$, $m=0.46m_0$ (where $m_0$ is free electron mass), $M=m_0$, $d=1$~nm, $l=2.5$~nm. 
    We also take $n_e=1.2\times 10^{12}$~cm$^{-2}$ and $n_c=5.0\times 10^{10}$~cm$^{-2}$. 
    }
    \label{Fig2}
\end{figure}
The SC critical temperature can be  estimated from Eq.~(\ref{EqGap11}) exploiting the condition $\Delta_{\lambda}(T_c^{\lambda})$=0. 
For 1b processes, it gives $T_c^{(1b)}=(\gamma/\pi)\Delta_{1b}(T=0)$, where $\gamma=\exp C_0$ with $C_0=0.577$ the Euler's constant (see, e.g.,~\cite{LandauLifshitz9}). 
The analytical estimation of $T_c^{(2b)}$ this way is cumbersome due to the presence of $N_q$-containing term in Eq.~(\ref{EqEffPot2b}). 

\section{Results and discussion}

Full temperature dependence of $\Delta_\lambda$ can be studied numerically using Eqs.~(\ref{EqEffPot1b})-(\ref{EqGap11}). 
Here, we account for the temperature dependence of the condensate density using the formula, which describes 2D BEC in a power-law trap~\cite{Bagnato1991},  $n_c(T)=n_c[1-(T/T_c^\textrm{BEC})^2]$, where $T_c^\textrm{BEC}$ is a critical temperature of the BEC formation. 
We take $T_c^\textrm{BEC}=100$~K in accordance with recent predictions~\cite{Fogler2014, WangNature2019}.
%
We also neglect the finite lifetime of bogolons, studied in works~\cite{Chung_2009, Kovalev2016} since in our case, the effective time of Cooper pair formation $\sim\Delta_{\lambda}^{-1}$ is smaller than the exciton scattering time on impurities $\tau$, $\Delta_{\lambda}\tau/(\xi_hk)^2\gg 1$.

Figure~\ref{Fig2} shows the comparison between the SC order parameters induced by 1b- and 2b-mediated pairings. 
%
%
%
\begin{figure}[!t]
    \centering
    \includegraphics[width=0.48\textwidth]{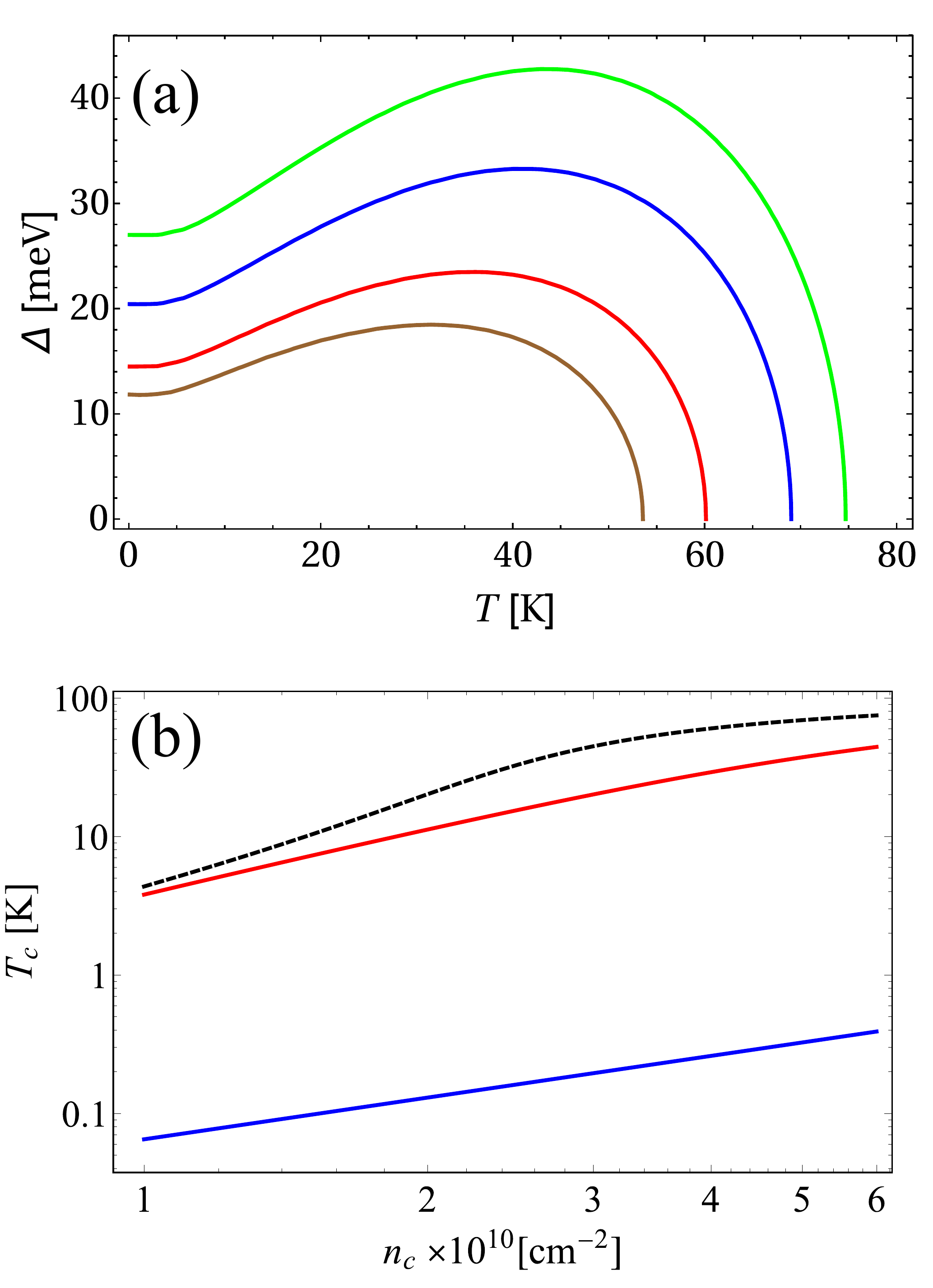}
    \caption{(a) Superconducting gap due to bogolon-pair-mediated processes as a function of temperature for different condensate densities: $n_c=3.5\times10^{10}$~cm$^{-2}$ (brown),
    $n_c=4.0\times10^{10}$~cm$^{-2}$ (red),
    $n_c=5.0\times10^{10}$~cm$^{-2}$ (blue),  and
    $n_c=6.0\times10^{10}$~cm$^{-2}$ (green). 
    (b) Critical temperature as a function of condensate density for single-bogolon processes (blue), two-bogolon processes without the $N_q$-containing term in Eq.~\eqref{EqEffPot2b} (red), and two-bogolon processes with the $N_q$-containing term (black dashed). 
    We used $n_e=1.0\times10^{12}$~cm$^{-2}$. All other parameters are the same as in Fig.~\ref{Fig2}. 
    }
    \label{Fig3}
\end{figure}
At the same condensate density $n_c$ and concentration of electrons in the 2DEG $n_e$, 2b-induced gap $\Delta_{2b}(T)$ is bigger than $\Delta_{1b}(T)$. 
This drastic difference between them is caused by the ratio of two effective electron-electron pairing potentials, $V_{1b}/V_{2b}\sim (\xi_h k_F)(n_c \xi_h^2)\ll 1$.
Moreover, the finite-temperature correction to the 2b-mediated pairing potential in Eq.~(\ref{EqEffPot2b}) leads to dramatic enhancement of the SC gap with the increase of temperature. 
As a result, 2b-induced order parameter reveals a pronounced non-monotonous temperature dependence.
We want to note, that non-monotonous dependence of the order parameter due to two-acoustic phonon-mediated pairing has been theoretically investigated in three-dimensional multi-band superconductors.
There, however, the two-phonon processes were considered as a second-order perturbation~\cite{Enaki_2002} giving a contribution in the absence of single-phonon processes. 
In our case, 2b pairing belongs to the same order of the perturbation theory as 1b pairing [see Eqs.~\eqref{EqEffPot1b} and~\eqref{EqEffPot2b}], as it will be discussed below.

We should also address the issue of Coulomb repulsion between electrons in 2DEG.
A standard calculation~\cite{mcmillan} gives the following renormalization of the coupling constant: $\tilde{V}_{\lambda}(p_F)\rightarrow V_{\lambda}(p_F)- V_C'$, where $V_C'=V_C/[1+\nu_0 V_C \log(\mu/\omega_b)]$ with $V_C$ the momentum-averaged Coulomb potential~\cite{RefEfetov}. 
Using the same parameters as in Fig.~\ref{Fig2}, we estimate $\nu_0V_C'\approx0.2$, while we consider $\nu_0V_{2b}$ in the range 0.4-1 (along the text).

It should also be noted, that our approach is valid in the \textit{weak electron-bogolon coupling regime} where the BCS theory is applicable~\cite{RefBookParks, Mahan}. 
It corresponds to $\nu_0 V_{2b}(p_F)<1$. 
Thus we only use $\nu_0 V_{2b}(p_F)$ in the range 0.4-1, where unity corresponds to a provisional  boundary, where the weak coupling regime breaks and a more sophisticated strong-coupling treatment within the Eliashberg equations approach is required~\cite{RefEliashberg1, RefEliashberg2, RefNambu, RefEfetov}.
However, we leave it beyond the scope of this article.

Figure~\ref{Fig3} shows the dependence of the 2b-mediated gap and the critical temperature on the condensate density.
As it follows from Eq.~\eqref{Gap2bog} (and Eq.~\eqref{Gap1bog} for 1b processes), both $\Delta$ and $T_c$ 
grow with the increase of $n_c$ (via the sound velocity $s$) or decrease of $n_e$ (via the Fermi wave vector $p_F$ in the exponential factor in $g_{p_F}$). 
A naive idea which comes to mind is to start increasing $n_c$ up to the maximal experimentally achievable values and decreasing $n_e$ while possible.
However, the applicability of the BCS theory imposes an additional requirement: $n_e/n_c> d/a^{el}_B$, where $a_B^{el}=\pi \epsilon_0\epsilon\hbar^2/m e_0^2$ is the Bohr radius of electrons in 2DEG. 
{\color{black} 
Meanwhile, considering only bogolons with a linear spectrum dictates another requirement: $k_F \xi < 1$, that gives the  condition $n_e/n_c<d/a_B^{ex}$, where $a_B^{ex}=\pi \epsilon_0\epsilon\hbar^2/M e_0^2$ is the Bohr radius of exciton. 
It results in a condition imposed on the effective masses: the effective electron mass in 2DEG should be smaller than the mass of the indirect exciton. 
The optimal relation between $n_e$ and $n_c$ is $n_e/n_c\sim C_1 \pi \epsilon_0\epsilon\hbar^2/m_0 e_0^2$, where $C_1$ is a numerical constant and $m_0$ is a free electron mass.
}
\begin{figure}[!b]
\includegraphics[width=0.49\textwidth]{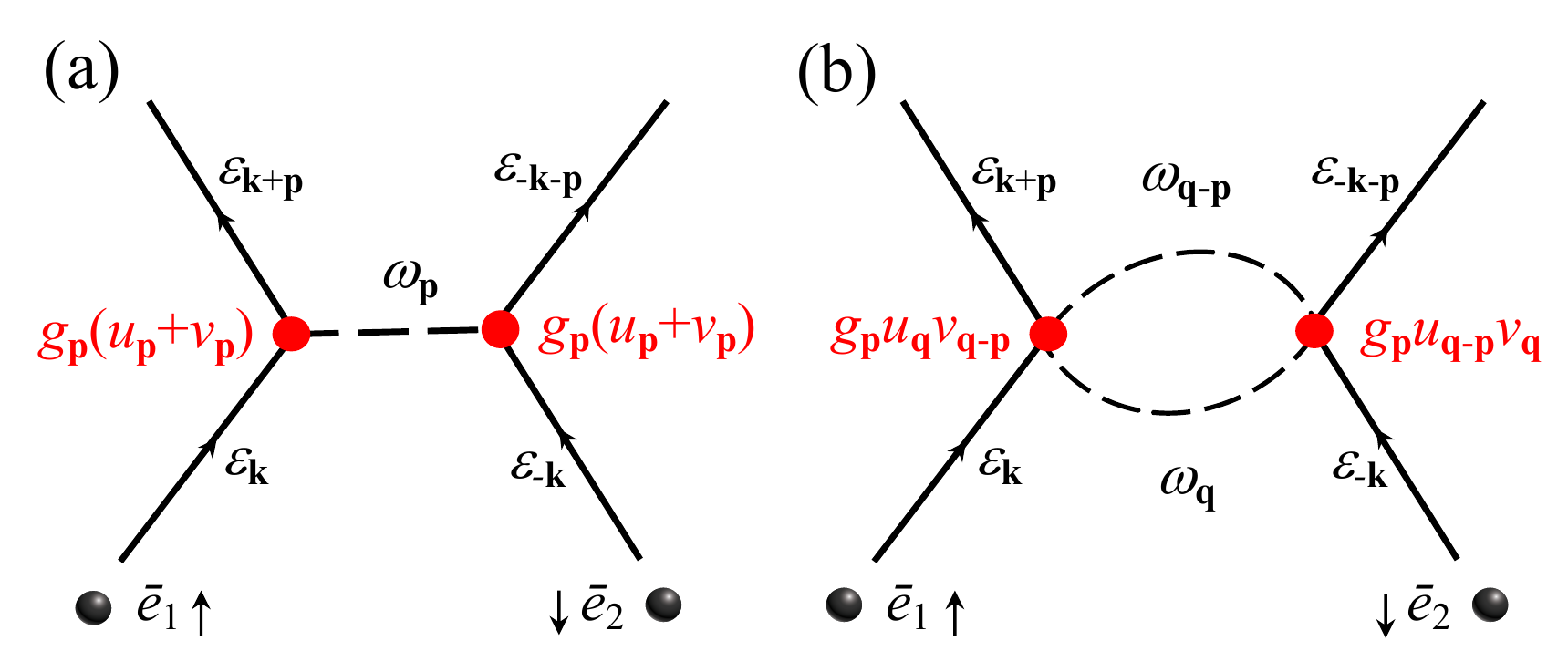}
\caption{Effective Feynman diagrams
of single-bogolon-mediated (a) and bogolon-pair-mediated (b) electron pairing.}
\label{Fig4}
\end{figure}
%
%
%


Why is 2b superconductivity stronger than 1b? {\color{black} The electron-single bogolon and electron-bogolon pair interactions are processes of the same order with respect to the electron-exciton interaction strength $g_p$ due to the properties of weakly interacting Bose gas at low temperature. 
The full density of the Bose gas consists of three parts: i) the condensate density $ n_c$, ii) density of excitations above the condensate  $\varphi^{\dag}_{\bf R}\varphi_{\bf R}$, and iii) the ``mixed density'' $\sqrt{n_c}(\varphi^{\dag}_{\bf R}+\varphi_{\bf R})$. 
This last term here does not conserve the number of Bose-particles in a given quantum state and usually gives small contribution to different physical processes, such as electron scattering, since only the non-diagonal matrix elements of this operator are nonzero, see Eq.~(\ref{EqTwoHam11}).} 

To understand the microscopic origin of this phenomenon,
in Fig.~\ref{Fig4} we show the Feynman diagrams, corresponding to 1b and 2b pairings, as it follows from the Schrieffer-Wolff transformation (see Supplemental Material~\cite{SMBG}). 
 The matrix elements of the electron-boson interaction $g_p$ are multiplied by the Bogoliubov coefficients. In the 1b case, it is the sum $(u_\mathbf{p}+v_{-\mathbf{p}})$, while in the 2b case a product of the kind $u_\mathbf{q}v_{\mathbf{q}-\mathbf{p}}$.
We see, that the key reason of suppression of the 1b processes is that there emerges a small factor $(u_\mathbf{p}+v_{-\mathbf{p}})\sim(p\xi_h)^2\ll 1$~\cite{Villegas2019}. 
Indeed, both $|u_\mathbf{p}|,~|v_\mathbf{p}|\gg1$, and they have opposite signs, thus negating each other in the sum.
It can be looked at as a destructive interference of waves corresponding to $b_\mathbf{p}$ and $b^\dagger_\mathbf{-p}$.
There is no such self-cancellation in the 2b matrix elements since $u_\textbf{p}v_\textbf{p}\sim (p\xi_h)^{-1}\gg 1$ (instead of $u_\mathbf{p}+v_{-\mathbf{p}}$). 
Here we can also recall the acoustic phonons, where such a cancellation effect does not take place, and hence the single-phonon scattering prevails over the two-phonon one, and thus the latter can be usually neglected. 
However, the physics in question is general and might be relevant to other proximity effects of the BEC phase.
We want to mention also, that the processes involving three and more bogolons belong to the higher-order perturbation theory {\color{black} with respect to the electron-exciton interaction $g_p$} and can be disregarded, as it has been discussed in~\cite{RefCapture}.

We note, that performing the calculations and evaluating the gap and $T_c$, we assumed that the electron gas is degenerate at given $n_e$ and temperature.
We have to also note, that the approach discussed in this article is only valid as long as $n_c$ is macroscopically large ($n_c\gtrsim 10^8$ cm$^{-2}$). 
Only under this condition, we can treat the bogolon dispersion as linear and use the mean field approach and the Bogoliubov transformations.

Certainly, SC $T_c$ should be smaller than $T_c^\textrm{BEC}$.
In GaAs-based excitonic structures, $T_c^\textrm{BEC}\sim 1-7$~K~\cite{Butov:2003aa} and it
is predicted to reach $\sim 100$~K or more in MoS$_2$~\cite{Fogler2014}, which finds its experimental signatures~\cite{WangNature2019}.
If the temperature is above the critical one, there is no BEC but electrons are still coupled with excitons via Coulomb forces. However, we believe that in this case Bose gas-mediated superconductivity is strongly suppressed~\cite{[ {The fundamental obstacle of the exciton-induced superconductivity can be associated with the parabolic dispersion of excitons, which velocity is much higher than the sound velocity of phonons or bogolons. As a result, electrons would attract each other at shorter distances, as compared with bogolon-mediated attraction. Then, short-size Cooper pairs should be strongly affected by the repulsive Coulomb interaction}]C2}.

Usually, the conventional phonon-mediated superconductivity is explained the following quantitative way: one electron moving along the crystal polarizes the media due to the Coulomb interaction between this electron and the nuclei, and then another electron (moving with the opposite or close-to-opposite momentum to the first electron) feels this polarization of the media, and by that the electrons effectively couple with each other. In our case, the ions of the crystal lattice are replaced by indirect excitons. 
And here, the mechanism of electron-electron pairing is similar qualitatively but quantitatively different: instead of the deformation potential, one deals with the direct Coulomb interaction between electrons and excitons, which can be treated as dipoles. 
Thus, the effective matrix elements of this interaction are different. 
As the result, one electron disturbs the excitonic media in BEC, while another one (with opposite momentum) feels the polarization, and the SC pairing might occur.


\section{Conclusions}

We have studied electron pairing in a two-dimensional electron gas in the vicinity of a two-dimensional Bose-Einstein condensate, taking a condensed dipolar exciton gas as an example. 
We have found that the bogolon-pair-mediated electron interaction turns out to be the dominant mechanism of pairing in hybrid systems, giving large superconducting gap and  critical temperatures of superconducting transition up to 80~K.
The effect is twofold. 
First, the bogolon-pair-induced gap is bigger than the single-bogolon one even at zero temperature due to the structure and magnitudes of the matrix elements of electron interaction.
Second, we predict that, in contrast to single-bogolon-mediated processes, two-bogolon electron pairing potential acquires an additional temperature-dependent term, associated with the increase of the number of thermally activated bogolons with temperature. 
As a consequence, such term leads to non-monotonous temperature characteristics of the superconducting gap and a considerable increase of $T_c$. 
We expect this exotic feature to be observable experimentally.
Moreover, instead of indirect excitons, one can employ microcavity exciton polaritons, where the BEC is reported to exist up to the room temperature~\cite{Lerario:2017aa}, or other bosons.


\section*{Acknowledgements}
We thank I.~Vakulchyk and I.~Krive for useful discussions.
We have been supported by the Institute for Basic Science in Korea (Project No.~IBS-R024-D1) and the Ministry of Science and Higher Education of the Russian Federation (Project No.~075-15-2020-797 (13.1902.21.0024)).


\bibliography{library}

\begin{thebibliography}{54}%
\makeatletter
\providecommand \@ifxundefined [1]{%
 \@ifx{#1\undefined}
}%
\providecommand \@ifnum [1]{%
 \ifnum #1\expandafter \@firstoftwo
 \else \expandafter \@secondoftwo
 \fi
}%
\providecommand \@ifx [1]{%
 \ifx #1\expandafter \@firstoftwo
 \else \expandafter \@secondoftwo
 \fi
}%
\providecommand \natexlab [1]{#1}%
\providecommand \enquote  [1]{``#1''}%
\providecommand \bibnamefont  [1]{#1}%
\providecommand \bibfnamefont [1]{#1}%
\providecommand \citenamefont [1]{#1}%
\providecommand \href@noop [0]{\@secondoftwo}%
\providecommand \href [0]{\begingroup \@sanitize@url \@href}%
\providecommand \@href[1]{\@@startlink{#1}\@@href}%
\providecommand \@@href[1]{\endgroup#1\@@endlink}%
\providecommand \@sanitize@url [0]{\catcode `\\12\catcode `\$12\catcode
  `\&12\catcode `\#12\catcode `\^12\catcode `\_12\catcode `\%12\relax}%
\providecommand \@@startlink[1]{}%
\providecommand \@@endlink[0]{}%
\providecommand \url  [0]{\begingroup\@sanitize@url \@url }%
\providecommand \@url [1]{\endgroup\@href {#1}{\urlprefix }}%
\providecommand \urlprefix  [0]{URL }%
\providecommand \Eprint [0]{\href }%
\providecommand \doibase [0]{http://dx.doi.org/}%
\providecommand \selectlanguage [0]{\@gobble}%
\providecommand \bibinfo  [0]{\@secondoftwo}%
\providecommand \bibfield  [0]{\@secondoftwo}%
\providecommand \translation [1]{[#1]}%
\providecommand \BibitemOpen [0]{}%
\providecommand \bibitemStop [0]{}%
\providecommand \bibitemNoStop [0]{.\EOS\space}%
\providecommand \EOS [0]{\spacefactor3000\relax}%
\providecommand \BibitemShut  [1]{\csname bibitem#1\endcsname}%
\let\auto@bib@innerbib\@empty
\bibitem [{\citenamefont {Bardeen}\ \emph
  {et~al.}(1957{\natexlab{a}})\citenamefont {Bardeen}, \citenamefont {Cooper},\
  and\ \citenamefont {Schrieffer}}]{RefMainBCS}%
  \BibitemOpen
  \bibfield  {author} {\bibinfo {author} {\bibfnamefont {J.}~\bibnamefont
  {Bardeen}}, \bibinfo {author} {\bibfnamefont {L.~N.}\ \bibnamefont {Cooper}},
  \ and\ \bibinfo {author} {\bibfnamefont {J.~R.}\ \bibnamefont {Schrieffer}},\
  }\href {\doibase 10.1103/PhysRev.106.162} {\bibfield  {journal} {\bibinfo
  {journal} {Phys. Rev.}\ }\textbf {\bibinfo {volume} {106}},\ \bibinfo {pages}
  {162} (\bibinfo {year} {1957}{\natexlab{a}})}\BibitemShut {NoStop}%
\bibitem [{\citenamefont {Bardeen}\ \emph
  {et~al.}(1957{\natexlab{b}})\citenamefont {Bardeen}, \citenamefont {Cooper},\
  and\ \citenamefont {Schrieffer}}]{PhysRev.108.1175}%
  \BibitemOpen
  \bibfield  {author} {\bibinfo {author} {\bibfnamefont {J.}~\bibnamefont
  {Bardeen}}, \bibinfo {author} {\bibfnamefont {L.~N.}\ \bibnamefont {Cooper}},
  \ and\ \bibinfo {author} {\bibfnamefont {J.~R.}\ \bibnamefont {Schrieffer}},\
  }\href {\doibase 10.1103/PhysRev.108.1175} {\bibfield  {journal} {\bibinfo
  {journal} {Phys. Rev.}\ }\textbf {\bibinfo {volume} {108}},\ \bibinfo {pages}
  {1175} (\bibinfo {year} {1957}{\natexlab{b}})}\BibitemShut {NoStop}%
\bibitem [{\citenamefont {Bednorz}\ and\ \citenamefont
  {M\"uller}(1986)}]{Bednorz1986}%
  \BibitemOpen
  \bibfield  {author} {\bibinfo {author} {\bibfnamefont {J.~G.}\ \bibnamefont
  {Bednorz}}\ and\ \bibinfo {author} {\bibfnamefont {K.~A.}\ \bibnamefont
  {M\"uller}},\ }\href {\doibase 10.1007/BF01303701} {\bibfield  {journal}
  {\bibinfo  {journal} {Z. Physik B - Cond. Mat.}\ }\textbf {\bibinfo {volume}
  {64}},\ \bibinfo {pages} {189} (\bibinfo {year} {1986})}\BibitemShut
  {NoStop}%
\bibitem [{\citenamefont {Gr\"uner}(1988)}]{RevModPhys.60.1129}%
  \BibitemOpen
  \bibfield  {author} {\bibinfo {author} {\bibfnamefont {G.}~\bibnamefont
  {Gr\"uner}},\ }\href {\doibase 10.1103/RevModPhys.60.1129} {\bibfield
  {journal} {\bibinfo  {journal} {Rev. Mod. Phys.}\ }\textbf {\bibinfo {volume}
  {60}},\ \bibinfo {pages} {1129} (\bibinfo {year} {1988})}\BibitemShut
  {NoStop}%
\bibitem [{\citenamefont {Cao}\ \emph {et~al.}(2018)\citenamefont {Cao},
  \citenamefont {Fatemi}, \citenamefont {Fang}, \citenamefont {Watanabe},
  \citenamefont {Taniguchi}, \citenamefont {Kaxiras},\ and\ \citenamefont
  {Jarillo-Herrero}}]{NatureBilayer}%
  \BibitemOpen
  \bibfield  {author} {\bibinfo {author} {\bibfnamefont {Y.}~\bibnamefont
  {Cao}}, \bibinfo {author} {\bibfnamefont {V.}~\bibnamefont {Fatemi}},
  \bibinfo {author} {\bibfnamefont {S.}~\bibnamefont {Fang}}, \bibinfo {author}
  {\bibfnamefont {K.}~\bibnamefont {Watanabe}}, \bibinfo {author}
  {\bibfnamefont {T.}~\bibnamefont {Taniguchi}}, \bibinfo {author}
  {\bibfnamefont {E.}~\bibnamefont {Kaxiras}}, \ and\ \bibinfo {author}
  {\bibfnamefont {P.}~\bibnamefont {Jarillo-Herrero}},\ }\href {\doibase
  10.1038/nature26160} {\bibfield  {journal} {\bibinfo  {journal} {Nature}\
  }\textbf {\bibinfo {volume} {556}},\ \bibinfo {pages} {43} (\bibinfo {year}
  {2018})}\BibitemShut {NoStop}%
\bibitem [{\citenamefont {Somayazulu}\ \emph {et~al.}(2019)\citenamefont
  {Somayazulu}, \citenamefont {Ahart}, \citenamefont {Mishra}, \citenamefont
  {Geballe}, \citenamefont {Baldini}, \citenamefont {Meng}, \citenamefont
  {Struzhkin},\ and\ \citenamefont {Hemley}}]{PhysRevLett.122.027001}%
  \BibitemOpen
  \bibfield  {author} {\bibinfo {author} {\bibfnamefont {M.}~\bibnamefont
  {Somayazulu}}, \bibinfo {author} {\bibfnamefont {M.}~\bibnamefont {Ahart}},
  \bibinfo {author} {\bibfnamefont {A.~K.}\ \bibnamefont {Mishra}}, \bibinfo
  {author} {\bibfnamefont {Z.~M.}\ \bibnamefont {Geballe}}, \bibinfo {author}
  {\bibfnamefont {M.}~\bibnamefont {Baldini}}, \bibinfo {author} {\bibfnamefont
  {Y.}~\bibnamefont {Meng}}, \bibinfo {author} {\bibfnamefont {V.~V.}\
  \bibnamefont {Struzhkin}}, \ and\ \bibinfo {author} {\bibfnamefont {R.~J.}\
  \bibnamefont {Hemley}},\ }\href {\doibase 10.1103/PhysRevLett.122.027001}
  {\bibfield  {journal} {\bibinfo  {journal} {Phys. Rev. Lett.}\ }\textbf
  {\bibinfo {volume} {122}},\ \bibinfo {pages} {027001} (\bibinfo {year}
  {2019})}\BibitemShut {NoStop}%
\bibitem [{\citenamefont {Frasca}\ and\ \citenamefont
  {Charbon}(2019)}]{Frasca2019}%
  \BibitemOpen
  \bibfield  {author} {\bibinfo {author} {\bibfnamefont {S.}~\bibnamefont
  {Frasca}}\ and\ \bibinfo {author} {\bibfnamefont {E.}~\bibnamefont
  {Charbon}},\ }\href {\doibase 10.1038/s41928-019-0319-x} {\bibfield
  {journal} {\bibinfo  {journal} {Nature Electron.}\ }\textbf {\bibinfo
  {volume} {2}},\ \bibinfo {pages} {433} (\bibinfo {year} {2019})}\BibitemShut
  {NoStop}%
\bibitem [{\citenamefont {Burkard}\ \emph {et~al.}(2020)\citenamefont
  {Burkard}, \citenamefont {Gullans}, \citenamefont {Mi},\ and\ \citenamefont
  {Petta}}]{Burkard2020}%
  \BibitemOpen
  \bibfield  {author} {\bibinfo {author} {\bibfnamefont {G.}~\bibnamefont
  {Burkard}}, \bibinfo {author} {\bibfnamefont {M.~J.}\ \bibnamefont
  {Gullans}}, \bibinfo {author} {\bibfnamefont {X.}~\bibnamefont {Mi}}, \ and\
  \bibinfo {author} {\bibfnamefont {J.~R.~P.}\ \bibnamefont {Petta}},\ }\href
  {\doibase 10.1038/s42254-019-0135-2} {\bibfield  {journal} {\bibinfo
  {journal} {Nature Rev. Phys.}\ }\textbf {\bibinfo {volume} {2}},\ \bibinfo
  {pages} {129} (\bibinfo {year} {2020})}\BibitemShut {NoStop}%
\bibitem [{\citenamefont {Villegas}\ \emph {et~al.}(2020)\citenamefont
  {Villegas}, \citenamefont {Kusmartsev}, \citenamefont {Luo},\ and\
  \citenamefont {Savenko}}]{PhysRevLett.124.087701}%
  \BibitemOpen
  \bibfield  {author} {\bibinfo {author} {\bibfnamefont {K.~H.~A.}\
  \bibnamefont {Villegas}}, \bibinfo {author} {\bibfnamefont {F.~V.}\
  \bibnamefont {Kusmartsev}}, \bibinfo {author} {\bibfnamefont
  {Y.}~\bibnamefont {Luo}}, \ and\ \bibinfo {author} {\bibfnamefont {I.~G.}\
  \bibnamefont {Savenko}},\ }\href {\doibase 10.1103/PhysRevLett.124.087701}
  {\bibfield  {journal} {\bibinfo  {journal} {Phys. Rev. Lett.}\ }\textbf
  {\bibinfo {volume} {124}},\ \bibinfo {pages} {087701} (\bibinfo {year}
  {2020})}\BibitemShut {NoStop}%
\bibitem [{\citenamefont {Saito}\ \emph {et~al.}(2016)\citenamefont {Saito},
  \citenamefont {Nakamura}, \citenamefont {Bahramy}, \citenamefont {Kohama},
  \citenamefont {Ye}, \citenamefont {Kasahara}, \citenamefont {Nakagawa},
  \citenamefont {Onga}, \citenamefont {Tokunaga}, \citenamefont {Nojima},
  \citenamefont {Yanase},\ and\ \citenamefont {Iwasa}}]{Saito2016}%
  \BibitemOpen
  \bibfield  {author} {\bibinfo {author} {\bibfnamefont {Y.}~\bibnamefont
  {Saito}}, \bibinfo {author} {\bibfnamefont {Y.}~\bibnamefont {Nakamura}},
  \bibinfo {author} {\bibfnamefont {M.~S.}\ \bibnamefont {Bahramy}}, \bibinfo
  {author} {\bibfnamefont {Y.}~\bibnamefont {Kohama}}, \bibinfo {author}
  {\bibfnamefont {J.}~\bibnamefont {Ye}}, \bibinfo {author} {\bibfnamefont
  {Y.}~\bibnamefont {Kasahara}}, \bibinfo {author} {\bibfnamefont
  {Y.}~\bibnamefont {Nakagawa}}, \bibinfo {author} {\bibfnamefont
  {M.}~\bibnamefont {Onga}}, \bibinfo {author} {\bibfnamefont {M.}~\bibnamefont
  {Tokunaga}}, \bibinfo {author} {\bibfnamefont {T.}~\bibnamefont {Nojima}},
  \bibinfo {author} {\bibfnamefont {Y.}~\bibnamefont {Yanase}}, \ and\ \bibinfo
  {author} {\bibfnamefont {Y.}~\bibnamefont {Iwasa}},\ }\href {\doibase
  10.1038/nphys3580} {\bibfield  {journal} {\bibinfo  {journal} {Nature Phys.}\
  }\textbf {\bibinfo {volume} {12}},\ \bibinfo {pages} {144} (\bibinfo {year}
  {2016})}\BibitemShut {NoStop}%
\bibitem [{\citenamefont {Ge}\ \emph {et~al.}(2015)\citenamefont {Ge},
  \citenamefont {Liu}, \citenamefont {Liu}, \citenamefont {Gao}, \citenamefont
  {Qian}, \citenamefont {Xue}, \citenamefont {Liu},\ and\ \citenamefont
  {Jia}}]{Ge2015}%
  \BibitemOpen
  \bibfield  {author} {\bibinfo {author} {\bibfnamefont {J.-F.}\ \bibnamefont
  {Ge}}, \bibinfo {author} {\bibfnamefont {Z.-L.}\ \bibnamefont {Liu}},
  \bibinfo {author} {\bibfnamefont {C.}~\bibnamefont {Liu}}, \bibinfo {author}
  {\bibfnamefont {C.-L.}\ \bibnamefont {Gao}}, \bibinfo {author} {\bibfnamefont
  {D.}~\bibnamefont {Qian}}, \bibinfo {author} {\bibfnamefont {Q.-K.}\
  \bibnamefont {Xue}}, \bibinfo {author} {\bibfnamefont {Y.}~\bibnamefont
  {Liu}}, \ and\ \bibinfo {author} {\bibfnamefont {J.-F.}\ \bibnamefont
  {Jia}},\ }\href {\doibase 10.1038/nmat4153} {\bibfield  {journal} {\bibinfo
  {journal} {Nature Mat.}\ }\textbf {\bibinfo {volume} {14}},\ \bibinfo {pages}
  {285} (\bibinfo {year} {2015})}\BibitemShut {NoStop}%
\bibitem [{\citenamefont {Uchihashi}(2016)}]{Uchihashi2016}%
  \BibitemOpen
  \bibfield  {author} {\bibinfo {author} {\bibfnamefont {T.}~\bibnamefont
  {Uchihashi}},\ }\href {\doibase 10.1088/0953-2048/30/1/013002} {\bibfield
  {journal} {\bibinfo  {journal} {Supercond. Sci. Tech.}\ }\textbf {\bibinfo
  {volume} {30}},\ \bibinfo {pages} {013002} (\bibinfo {year}
  {2016})}\BibitemShut {NoStop}%
\bibitem [{\citenamefont {Little}(1964)}]{little}%
  \BibitemOpen
  \bibfield  {author} {\bibinfo {author} {\bibfnamefont {W.~A.}\ \bibnamefont
  {Little}},\ }\href {\doibase 10.1103/PhysRev.134.A1416} {\bibfield  {journal}
  {\bibinfo  {journal} {Phys. Rev.}\ }\textbf {\bibinfo {volume} {134}},\
  \bibinfo {pages} {A1416} (\bibinfo {year} {1964})}\BibitemShut {NoStop}%
\bibitem [{\citenamefont {V.~L.~Ginzburg}(1972)}]{ginzburg}%
  \BibitemOpen
  \bibfield  {author} {\bibinfo {author} {\bibfnamefont {D.~A.~K.}\
  \bibnamefont {V.~L.~Ginzburg}},\ }\href@noop {} {\bibfield  {journal}
  {\bibinfo  {journal} {Phys. Reports}\ }\textbf {\bibinfo {volume} {4}},\
  \bibinfo {pages} {343} (\bibinfo {year} {1972})}\BibitemShut {NoStop}%
\bibitem [{\citenamefont {Allender}\ \emph {et~al.}(1973)\citenamefont
  {Allender}, \citenamefont {Bray},\ and\ \citenamefont
  {Bardeen}}]{PhysRevB.7.1020}%
  \BibitemOpen
  \bibfield  {author} {\bibinfo {author} {\bibfnamefont {D.}~\bibnamefont
  {Allender}}, \bibinfo {author} {\bibfnamefont {J.}~\bibnamefont {Bray}}, \
  and\ \bibinfo {author} {\bibfnamefont {J.}~\bibnamefont {Bardeen}},\ }\href
  {\doibase 10.1103/PhysRevB.7.1020} {\bibfield  {journal} {\bibinfo  {journal}
  {Phys. Rev. B}\ }\textbf {\bibinfo {volume} {7}},\ \bibinfo {pages} {1020}
  (\bibinfo {year} {1973})}\BibitemShut {NoStop}%
\bibitem [{\citenamefont {Schlawin}\ \emph {et~al.}(2019)\citenamefont
  {Schlawin}, \citenamefont {Cavalleri},\ and\ \citenamefont
  {Jaksch}}]{PhysRevLett.122.133602}%
  \BibitemOpen
  \bibfield  {author} {\bibinfo {author} {\bibfnamefont {F.}~\bibnamefont
  {Schlawin}}, \bibinfo {author} {\bibfnamefont {A.}~\bibnamefont {Cavalleri}},
  \ and\ \bibinfo {author} {\bibfnamefont {D.}~\bibnamefont {Jaksch}},\ }\href
  {\doibase 10.1103/PhysRevLett.122.133602} {\bibfield  {journal} {\bibinfo
  {journal} {Phys. Rev. Lett.}\ }\textbf {\bibinfo {volume} {122}},\ \bibinfo
  {pages} {133602} (\bibinfo {year} {2019})}\BibitemShut {NoStop}%
\bibitem [{\citenamefont {Imamo{\u g}lu}\ \emph {et~al.}(1996)\citenamefont
  {Imamo{\u g}lu}, \citenamefont {Ram}, \citenamefont {Pau},\ and\
  \citenamefont {Yamamoto}}]{PhysRevA.53.4250}%
  \BibitemOpen
  \bibfield  {author} {\bibinfo {author} {\bibfnamefont {A.}~\bibnamefont
  {Imamo{\u g}lu}}, \bibinfo {author} {\bibfnamefont {R.~J.}\ \bibnamefont
  {Ram}}, \bibinfo {author} {\bibfnamefont {S.}~\bibnamefont {Pau}}, \ and\
  \bibinfo {author} {\bibfnamefont {Y.}~\bibnamefont {Yamamoto}},\ }\href
  {\doibase 10.1103/PhysRevA.53.4250} {\bibfield  {journal} {\bibinfo
  {journal} {Phys. Rev. A}\ }\textbf {\bibinfo {volume} {53}},\ \bibinfo
  {pages} {4250} (\bibinfo {year} {1996})}\BibitemShut {NoStop}%
\bibitem [{\citenamefont {Yu. E.~Lozovik}(1976)}]{lozovik}%
  \BibitemOpen
  \bibfield  {author} {\bibinfo {author} {\bibfnamefont {V.~I.~Y.}\
  \bibnamefont {Yu. E.~Lozovik}},\ }\href@noop {} {\bibfield  {journal}
  {\bibinfo  {journal} {Sov. Phys. JETP}\ }\textbf {\bibinfo {volume} {44}},\
  \bibinfo {pages} {389} (\bibinfo {year} {1976})}\BibitemShut {NoStop}%
\bibitem [{\citenamefont {Fogler}\ \emph {et~al.}(2016)\citenamefont {Fogler},
  \citenamefont {Butov},\ and\ \citenamefont {Novoselov}}]{Fogler2014}%
  \BibitemOpen
  \bibfield  {author} {\bibinfo {author} {\bibfnamefont {M.~M.}\ \bibnamefont
  {Fogler}}, \bibinfo {author} {\bibfnamefont {L.}~\bibnamefont {Butov}}, \
  and\ \bibinfo {author} {\bibfnamefont {K.~S.}\ \bibnamefont {Novoselov}},\
  }\href {\doibase 10.1038/ncomms5555} {\bibfield  {journal} {\bibinfo
  {journal} {Nature Comm.}\ }\textbf {\bibinfo {volume} {5}},\ \bibinfo {pages}
  {4555} (\bibinfo {year} {2016})}\BibitemShut {NoStop}%
\bibitem [{\citenamefont {Wu}\ \emph {et~al.}(2015)\citenamefont {Wu},
  \citenamefont {Xue},\ and\ \citenamefont {MacDonald}}]{PhysRevB.92.165121}%
  \BibitemOpen
  \bibfield  {author} {\bibinfo {author} {\bibfnamefont {F.-C.}\ \bibnamefont
  {Wu}}, \bibinfo {author} {\bibfnamefont {F.}~\bibnamefont {Xue}}, \ and\
  \bibinfo {author} {\bibfnamefont {A.~H.}\ \bibnamefont {MacDonald}},\ }\href
  {\doibase 10.1103/PhysRevB.92.165121} {\bibfield  {journal} {\bibinfo
  {journal} {Phys. Rev. B}\ }\textbf {\bibinfo {volume} {92}},\ \bibinfo
  {pages} {165121} (\bibinfo {year} {2015})}\BibitemShut {NoStop}%
\bibitem [{\citenamefont {Berman}\ and\ \citenamefont
  {Kezerashvili}(2016)}]{PhysRevB.93.245410}%
  \BibitemOpen
  \bibfield  {author} {\bibinfo {author} {\bibfnamefont {O.~L.}\ \bibnamefont
  {Berman}}\ and\ \bibinfo {author} {\bibfnamefont {R.~Y.}\ \bibnamefont
  {Kezerashvili}},\ }\href {\doibase 10.1103/PhysRevB.93.245410} {\bibfield
  {journal} {\bibinfo  {journal} {Phys. Rev. B}\ }\textbf {\bibinfo {volume}
  {93}},\ \bibinfo {pages} {245410} (\bibinfo {year} {2016})}\BibitemShut
  {NoStop}%
\bibitem [{\citenamefont {Debnath}\ \emph {et~al.}(2017)\citenamefont
  {Debnath}, \citenamefont {Barlas}, \citenamefont {Wickramaratne},
  \citenamefont {Neupane},\ and\ \citenamefont {Lake}}]{PhysRevB.96.174504}%
  \BibitemOpen
  \bibfield  {author} {\bibinfo {author} {\bibfnamefont {B.}~\bibnamefont
  {Debnath}}, \bibinfo {author} {\bibfnamefont {Y.}~\bibnamefont {Barlas}},
  \bibinfo {author} {\bibfnamefont {D.}~\bibnamefont {Wickramaratne}}, \bibinfo
  {author} {\bibfnamefont {M.~R.}\ \bibnamefont {Neupane}}, \ and\ \bibinfo
  {author} {\bibfnamefont {R.~K.}\ \bibnamefont {Lake}},\ }\href {\doibase
  10.1103/PhysRevB.96.174504} {\bibfield  {journal} {\bibinfo  {journal} {Phys.
  Rev. B}\ }\textbf {\bibinfo {volume} {96}},\ \bibinfo {pages} {174504}
  (\bibinfo {year} {2017})}\BibitemShut {NoStop}%
\bibitem [{\citenamefont {Su}\ \emph {et~al.}(2020)\citenamefont {Su},
  \citenamefont {Ghosh}, \citenamefont {Wang}, \citenamefont {Liu},
  \citenamefont {Diederichs}, \citenamefont {Liew},\ and\ \citenamefont
  {Xiong}}]{Room1}%
  \BibitemOpen
  \bibfield  {author} {\bibinfo {author} {\bibfnamefont {R.}~\bibnamefont
  {Su}}, \bibinfo {author} {\bibfnamefont {S.}~\bibnamefont {Ghosh}}, \bibinfo
  {author} {\bibfnamefont {J.}~\bibnamefont {Wang}}, \bibinfo {author}
  {\bibfnamefont {S.}~\bibnamefont {Liu}}, \bibinfo {author} {\bibfnamefont
  {C.}~\bibnamefont {Diederichs}}, \bibinfo {author} {\bibfnamefont {T.~C.~H.}\
  \bibnamefont {Liew}}, \ and\ \bibinfo {author} {\bibfnamefont
  {Q.}~\bibnamefont {Xiong}},\ }\href {\doibase 10.1038/s41567-019-0764-5}
  {\bibfield  {journal} {\bibinfo  {journal} {Nature Phys.}\ }\textbf {\bibinfo
  {volume} {16}},\ \bibinfo {pages} {301–306} (\bibinfo {year}
  {2020})}\BibitemShut {NoStop}%
\bibitem [{\citenamefont {Plumhof}\ \emph {et~al.}(2014)\citenamefont
  {Plumhof}, \citenamefont {St\"oferle}, \citenamefont {Mai}, \citenamefont
  {Scherf},\ and\ \citenamefont {Mahrt}}]{Room2}%
  \BibitemOpen
  \bibfield  {author} {\bibinfo {author} {\bibfnamefont {J.~D.}\ \bibnamefont
  {Plumhof}}, \bibinfo {author} {\bibfnamefont {T.}~\bibnamefont {St\"oferle}},
  \bibinfo {author} {\bibfnamefont {L.}~\bibnamefont {Mai}}, \bibinfo {author}
  {\bibfnamefont {U.}~\bibnamefont {Scherf}}, \ and\ \bibinfo {author}
  {\bibfnamefont {R.~F.}\ \bibnamefont {Mahrt}},\ }\href {\doibase
  10.1038/nmat3825} {\bibfield  {journal} {\bibinfo  {journal} {Nature Mat.}\
  }\textbf {\bibinfo {volume} {13}},\ \bibinfo {pages} {247} (\bibinfo {year}
  {2014})}\BibitemShut {NoStop}%
\bibitem [{\citenamefont {Wang}\ \emph {et~al.}(2019)\citenamefont {Wang},
  \citenamefont {Rhodes}, \citenamefont {Watanabe}, \citenamefont {Taniguchi},
  \citenamefont {Hone}, \citenamefont {Shan},\ and\ \citenamefont
  {Mak}}]{WangNature2019}%
  \BibitemOpen
  \bibfield  {author} {\bibinfo {author} {\bibfnamefont {Z.}~\bibnamefont
  {Wang}}, \bibinfo {author} {\bibfnamefont {D.~A.}\ \bibnamefont {Rhodes}},
  \bibinfo {author} {\bibfnamefont {K.}~\bibnamefont {Watanabe}}, \bibinfo
  {author} {\bibfnamefont {T.}~\bibnamefont {Taniguchi}}, \bibinfo {author}
  {\bibfnamefont {J.~C.}\ \bibnamefont {Hone}}, \bibinfo {author}
  {\bibfnamefont {J.}~\bibnamefont {Shan}}, \ and\ \bibinfo {author}
  {\bibfnamefont {K.~F.}\ \bibnamefont {Mak}},\ }\href {\doibase
  10.1038/s41586-019-1591-7} {\bibfield  {journal} {\bibinfo  {journal}
  {Nature}\ }\textbf {\bibinfo {volume} {574}},\ \bibinfo {pages} {76}
  (\bibinfo {year} {2019})}\BibitemShut {NoStop}%
\bibitem [{\citenamefont {Zhang}\ \emph {et~al.}(2014)\citenamefont {Zhang},
  \citenamefont {Johnson}, \citenamefont {Hsu}, \citenamefont {Li},\ and\
  \citenamefont {Shih}}]{doi:10.1021/nl501133c}%
  \BibitemOpen
  \bibfield  {author} {\bibinfo {author} {\bibfnamefont {C.}~\bibnamefont
  {Zhang}}, \bibinfo {author} {\bibfnamefont {A.}~\bibnamefont {Johnson}},
  \bibinfo {author} {\bibfnamefont {C.-L.}\ \bibnamefont {Hsu}}, \bibinfo
  {author} {\bibfnamefont {L.-J.}\ \bibnamefont {Li}}, \ and\ \bibinfo {author}
  {\bibfnamefont {C.-K.}\ \bibnamefont {Shih}},\ }\href {\doibase
  10.1021/nl501133c} {\bibfield  {journal} {\bibinfo  {journal} {Nano Lett.}\
  }\textbf {\bibinfo {volume} {14}},\ \bibinfo {pages} {2443} (\bibinfo {year}
  {2014})}\BibitemShut {NoStop}%
\bibitem [{\citenamefont {Laussy}\ \emph {et~al.}(2010)\citenamefont {Laussy},
  \citenamefont {Kavokin},\ and\ \citenamefont {Shelykh}}]{Laussy:2010aa}%
  \BibitemOpen
  \bibfield  {author} {\bibinfo {author} {\bibfnamefont {F.~P.}\ \bibnamefont
  {Laussy}}, \bibinfo {author} {\bibfnamefont {A.~V.}\ \bibnamefont {Kavokin}},
  \ and\ \bibinfo {author} {\bibfnamefont {I.~A.}\ \bibnamefont {Shelykh}},\
  }\href {\doibase 10.1103/PhysRevLett.104.106402} {\bibfield  {journal}
  {\bibinfo  {journal} {Phys. Rev. Lett.}\ }\textbf {\bibinfo {volume} {104}},\
  \bibinfo {pages} {106402} (\bibinfo {year} {2010})}\BibitemShut {NoStop}%
\bibitem [{\citenamefont {Cotle{\c t}}\ \emph {et~al.}(2016)\citenamefont
  {Cotle{\c t}}, \citenamefont {Zeytino\ifmmode~\check{g}\else \v{g}\fi{}lu},
  \citenamefont {Sigrist}, \citenamefont {Demler},\ and\ \citenamefont
  {Imamo\ifmmode~\check{g}\else \v{g}\fi{}lu}}]{Cotleifmmode-telse-tfi:2016aa}%
  \BibitemOpen
  \bibfield  {author} {\bibinfo {author} {\bibfnamefont {O.}~\bibnamefont
  {Cotle{\c t}}}, \bibinfo {author} {\bibfnamefont {S.}~\bibnamefont
  {Zeytino\ifmmode~\check{g}\else \v{g}\fi{}lu}}, \bibinfo {author}
  {\bibfnamefont {M.}~\bibnamefont {Sigrist}}, \bibinfo {author} {\bibfnamefont
  {E.}~\bibnamefont {Demler}}, \ and\ \bibinfo {author} {\bibfnamefont
  {A.}~\bibnamefont {Imamo\ifmmode~\check{g}\else \v{g}\fi{}lu}},\ }\href
  {\doibase 10.1103/PhysRevB.93.054510} {\bibfield  {journal} {\bibinfo
  {journal} {Phys. Rev. B}\ }\textbf {\bibinfo {volume} {93}},\ \bibinfo
  {pages} {054510} (\bibinfo {year} {2016})}\BibitemShut {NoStop}%
\bibitem [{\citenamefont {Skopelitis}\ \emph {et~al.}(2018)\citenamefont
  {Skopelitis}, \citenamefont {Cherotchenko}, \citenamefont {Kavokin},\ and\
  \citenamefont {Posazhennikova}}]{Skopelitis:2018aa}%
  \BibitemOpen
  \bibfield  {author} {\bibinfo {author} {\bibfnamefont {P.}~\bibnamefont
  {Skopelitis}}, \bibinfo {author} {\bibfnamefont {E.~D.}\ \bibnamefont
  {Cherotchenko}}, \bibinfo {author} {\bibfnamefont {A.~V.}\ \bibnamefont
  {Kavokin}}, \ and\ \bibinfo {author} {\bibfnamefont {A.}~\bibnamefont
  {Posazhennikova}},\ }\href {\doibase 10.1103/PhysRevLett.120.107001}
  {\bibfield  {journal} {\bibinfo  {journal} {Phys. Rev. Lett.}\ }\textbf
  {\bibinfo {volume} {120}},\ \bibinfo {pages} {107001} (\bibinfo {year}
  {2018})}\BibitemShut {NoStop}%
\bibitem [{\citenamefont {Villegas}\ \emph {et~al.}(2019)\citenamefont
  {Villegas}, \citenamefont {Sun}, \citenamefont {Kovalev},\ and\ \citenamefont
  {Savenko}}]{Villegas2019}%
  \BibitemOpen
  \bibfield  {author} {\bibinfo {author} {\bibfnamefont {K.~H.~A.}\
  \bibnamefont {Villegas}}, \bibinfo {author} {\bibfnamefont {M.}~\bibnamefont
  {Sun}}, \bibinfo {author} {\bibfnamefont {V.~M.}\ \bibnamefont {Kovalev}}, \
  and\ \bibinfo {author} {\bibfnamefont {I.~G.}\ \bibnamefont {Savenko}},\
  }\href {\doibase 10.1103/PhysRevLett.123.095301} {\bibfield  {journal}
  {\bibinfo  {journal} {Phys. Rev. Lett.}\ }\textbf {\bibinfo {volume} {123}},\
  \bibinfo {pages} {095301} (\bibinfo {year} {2019})}\BibitemShut {NoStop}%
\bibitem [{\citenamefont {Butov}(2017)}]{Butov2017}%
  \BibitemOpen
  \bibfield  {author} {\bibinfo {author} {\bibfnamefont {L.}~\bibnamefont
  {Butov}},\ }\href {\doibase 10.1016/j.spmi.2016.12.035} {\bibfield  {journal}
  {\bibinfo  {journal} {Superlatt. and Microstr.}\ }\textbf {\bibinfo {volume}
  {108}},\ \bibinfo {pages} {2 } (\bibinfo {year} {2017})}\BibitemShut
  {NoStop}%
\bibitem [{\citenamefont {Boev}\ \emph {et~al.}(2016)\citenamefont {Boev},
  \citenamefont {Kovalev},\ and\ \citenamefont {Savenko}}]{Boev:2016aa}%
  \BibitemOpen
  \bibfield  {author} {\bibinfo {author} {\bibfnamefont {M.~V.}\ \bibnamefont
  {Boev}}, \bibinfo {author} {\bibfnamefont {V.~M.}\ \bibnamefont {Kovalev}}, \
  and\ \bibinfo {author} {\bibfnamefont {I.~G.}\ \bibnamefont {Savenko}},\
  }\href {\doibase 10.1103/PhysRevB.94.241408} {\bibfield  {journal} {\bibinfo
  {journal} {Phys. Rev. B}\ }\textbf {\bibinfo {volume} {94}},\ \bibinfo
  {pages} {241408} (\bibinfo {year} {2016})}\BibitemShut {NoStop}%
\bibitem [{\citenamefont {Matuszewski}\ \emph {et~al.}(2012)\citenamefont
  {Matuszewski}, \citenamefont {Taylor},\ and\ \citenamefont
  {Kavokin}}]{Matuszewski:2012aa}%
  \BibitemOpen
  \bibfield  {author} {\bibinfo {author} {\bibfnamefont {M.}~\bibnamefont
  {Matuszewski}}, \bibinfo {author} {\bibfnamefont {T.}~\bibnamefont {Taylor}},
  \ and\ \bibinfo {author} {\bibfnamefont {A.~V.}\ \bibnamefont {Kavokin}},\
  }\href {\doibase 10.1103/PhysRevLett.108.060401} {\bibfield  {journal}
  {\bibinfo  {journal} {Phys. Rev. Lett.}\ }\textbf {\bibinfo {volume} {108}},\
  \bibinfo {pages} {060401} (\bibinfo {year} {2012})}\BibitemShut {NoStop}%
\bibitem [{\citenamefont {Giorgini}(1998)}]{Giorgini:1998aa}%
  \BibitemOpen
  \bibfield  {author} {\bibinfo {author} {\bibfnamefont {S.}~\bibnamefont
  {Giorgini}},\ }\href {\doibase 10.1103/PhysRevA.57.2949} {\bibfield
  {journal} {\bibinfo  {journal} {Phys. Rev. A}\ }\textbf {\bibinfo {volume}
  {57}},\ \bibinfo {pages} {2949} (\bibinfo {year} {1998})}\BibitemShut
  {NoStop}%
\bibitem [{C1()}]{C1}%
  \BibitemOpen
  \href@noop {} {\ }\BibitemShut {NoStop}%
\bibitem [{\citenamefont {Mahan}(1990)}]{Mahan}%
  \BibitemOpen
  \bibfield  {author} {\bibinfo {author} {\bibfnamefont {G.~D.}\ \bibnamefont
  {Mahan}},\ }\href@noop {} {\emph {\bibinfo {title} {{Many-Particle
  Physics}}}}\ (\bibinfo  {publisher} {Plenum Press, New York and London},\
  \bibinfo {year} {1990})\BibitemShut {NoStop}%
\bibitem [{C4()}]{C4}%
  \BibitemOpen
  \href@noop {} {\ }\BibitemShut {NoStop}%
\bibitem [{SMB()}]{SMBG}%
  \BibitemOpen
  \href@noop {} {\ }\BibitemShut {NoStop}%
\bibitem [{\citenamefont {Hohenberg}(1967)}]{Hohenberg1967}%
  \BibitemOpen
  \bibfield  {author} {\bibinfo {author} {\bibfnamefont {P.~C.}\ \bibnamefont
  {Hohenberg}},\ }\href {\doibase 10.1103/PhysRev.158.383} {\bibfield
  {journal} {\bibinfo  {journal} {Phys. Rev.}\ }\textbf {\bibinfo {volume}
  {158}},\ \bibinfo {pages} {383} (\bibinfo {year} {1967})}\BibitemShut
  {NoStop}%
\bibitem [{\citenamefont {Bagnato}\ and\ \citenamefont
  {Kleppner}(1991)}]{Bagnato1991}%
  \BibitemOpen
  \bibfield  {author} {\bibinfo {author} {\bibfnamefont {V.}~\bibnamefont
  {Bagnato}}\ and\ \bibinfo {author} {\bibfnamefont {D.}~\bibnamefont
  {Kleppner}},\ }\href {\doibase 10.1103/PhysRevA.44.7439} {\bibfield
  {journal} {\bibinfo  {journal} {Phys. Rev. A}\ }\textbf {\bibinfo {volume}
  {44}},\ \bibinfo {pages} {7439} (\bibinfo {year} {1991})}\BibitemShut
  {NoStop}%
\bibitem [{\citenamefont {Lifshitz}\ and\ \citenamefont
  {Pitaevskii}(1980)}]{LandauLifshitz9}%
  \BibitemOpen
  \bibfield  {author} {\bibinfo {author} {\bibfnamefont {E.~M.}\ \bibnamefont
  {Lifshitz}}\ and\ \bibinfo {author} {\bibfnamefont {L.~P.}\ \bibnamefont
  {Pitaevskii}},\ }\href@noop {} {\emph {\bibinfo {title} {{Statistical
  Physics, Part 2: The Theory of Condensed State}}}},\ Vol.~\bibinfo {volume}
  {9}\ (\bibinfo  {publisher} {Butterworth-Heinemann},\ \bibinfo {year}
  {1980})\BibitemShut {NoStop}%
\bibitem [{\citenamefont {Chung}\ and\ \citenamefont
  {Bhattacherjee}(2009)}]{Chung_2009}%
  \BibitemOpen
  \bibfield  {author} {\bibinfo {author} {\bibfnamefont {M.-C.}\ \bibnamefont
  {Chung}}\ and\ \bibinfo {author} {\bibfnamefont {A.~B.}\ \bibnamefont
  {Bhattacherjee}},\ }\href {\doibase 10.1088/1367-2630/11/12/123012}
  {\bibfield  {journal} {\bibinfo  {journal} {New J. Phys.}\ }\textbf {\bibinfo
  {volume} {11}},\ \bibinfo {pages} {123012} (\bibinfo {year}
  {2009})}\BibitemShut {NoStop}%
\bibitem [{\citenamefont {Kovalev}\ and\ \citenamefont
  {Chaplik}(2016)}]{Kovalev2016}%
  \BibitemOpen
  \bibfield  {author} {\bibinfo {author} {\bibfnamefont {V.~M.}\ \bibnamefont
  {Kovalev}}\ and\ \bibinfo {author} {\bibfnamefont {A.~V.}\ \bibnamefont
  {Chaplik}},\ }\href {\doibase 10.1134/S1063776116030158} {\bibfield
  {journal} {\bibinfo  {journal} {JETP}\ }\textbf {\bibinfo {volume} {122}},\
  \bibinfo {pages} {499} (\bibinfo {year} {2016})}\BibitemShut {NoStop}%
\bibitem [{\citenamefont {Enaki}\ and\ \citenamefont
  {Eremeev}(2002)}]{Enaki_2002}%
  \BibitemOpen
  \bibfield  {author} {\bibinfo {author} {\bibfnamefont {N.~A.}\ \bibnamefont
  {Enaki}}\ and\ \bibinfo {author} {\bibfnamefont {V.~V.}\ \bibnamefont
  {Eremeev}},\ }\href {\doibase 10.1088/1367-2630/4/1/380} {\bibfield
  {journal} {\bibinfo  {journal} {New J. Phys.}\ }\textbf {\bibinfo {volume}
  {4}},\ \bibinfo {pages} {80} (\bibinfo {year} {2002})}\BibitemShut {NoStop}%
\bibitem [{\citenamefont {McMillan}(1968)}]{mcmillan}%
  \BibitemOpen
  \bibfield  {author} {\bibinfo {author} {\bibfnamefont {W.~L.}\ \bibnamefont
  {McMillan}},\ }\href {\doibase 10.1103/PhysRev.167.331} {\bibfield  {journal}
  {\bibinfo  {journal} {Phys. Rev.}\ }\textbf {\bibinfo {volume} {167}},\
  \bibinfo {pages} {331} (\bibinfo {year} {1968})}\BibitemShut {NoStop}%
\bibitem [{\citenamefont {Einenkel}\ and\ \citenamefont
  {Efetov}(2011)}]{RefEfetov}%
  \BibitemOpen
  \bibfield  {author} {\bibinfo {author} {\bibfnamefont {M.}~\bibnamefont
  {Einenkel}}\ and\ \bibinfo {author} {\bibfnamefont {K.~B.}\ \bibnamefont
  {Efetov}},\ }\href {\doibase 10.1103/PhysRevB.84.214508} {\bibfield
  {journal} {\bibinfo  {journal} {Phys. Rev. B}\ }\textbf {\bibinfo {volume}
  {84}},\ \bibinfo {pages} {214508} (\bibinfo {year} {2011})}\BibitemShut
  {NoStop}%
\bibitem [{\citenamefont {Parks}(1969)}]{RefBookParks}%
  \BibitemOpen
  \bibfield  {author} {\bibinfo {author} {\bibfnamefont {R.~D.}\ \bibnamefont
  {Parks}},\ }\href@noop {} {\emph {\bibinfo {title} {Superconductivity (in two
  volumes)}}}\ (\bibinfo  {publisher} {Marcel Dekker, Inc., New York},\
  \bibinfo {year} {1969})\BibitemShut {NoStop}%
\bibitem [{\citenamefont {Eliashberg}(1960)}]{RefEliashberg1}%
  \BibitemOpen
  \bibfield  {author} {\bibinfo {author} {\bibfnamefont {G.~M.}\ \bibnamefont
  {Eliashberg}},\ }\href
  {http://www.w2agz.com/Library/Classic\%20Papers\%20in\%20Superconductivity/Eliashberg,\%20e-p\%20Interactions\%20in\%20SCs,\%20Sov-Phys\%20JETP\%2011,\%20696\%20(1960).pdf}
  {\bibfield  {journal} {\bibinfo  {journal} {Sov. Phys. JETP}\ }\textbf
  {\bibinfo {volume} {11}},\ \bibinfo {pages} {696} (\bibinfo {year}
  {1960})}\BibitemShut {NoStop}%
\bibitem [{\citenamefont {Eliashberg}(1961)}]{RefEliashberg2}%
  \BibitemOpen
  \bibfield  {author} {\bibinfo {author} {\bibfnamefont {G.~M.}\ \bibnamefont
  {Eliashberg}},\ }\href {http://www.jetp.ac.ru/cgi-bin/dn/e_012_05_1000.pdf}
  {\bibfield  {journal} {\bibinfo  {journal} {Sov. Phys. JETP}\ }\textbf
  {\bibinfo {volume} {12}},\ \bibinfo {pages} {1000} (\bibinfo {year}
  {1961})}\BibitemShut {NoStop}%
\bibitem [{\citenamefont {Nambu}(1960)}]{RefNambu}%
  \BibitemOpen
  \bibfield  {author} {\bibinfo {author} {\bibfnamefont {Y.}~\bibnamefont
  {Nambu}},\ }\href {\doibase 10.1103/PhysRev.117.648} {\bibfield  {journal}
  {\bibinfo  {journal} {Phys. Rev.}\ }\textbf {\bibinfo {volume} {117}},\
  \bibinfo {pages} {648} (\bibinfo {year} {1960})}\BibitemShut {NoStop}%
\bibitem [{\citenamefont {Boev}\ \emph {et~al.}(2018)\citenamefont {Boev},
  \citenamefont {Kovalev},\ and\ \citenamefont {Savenko}}]{RefCapture}%
  \BibitemOpen
  \bibfield  {author} {\bibinfo {author} {\bibfnamefont {M.~V.}\ \bibnamefont
  {Boev}}, \bibinfo {author} {\bibfnamefont {V.~M.}\ \bibnamefont {Kovalev}}, \
  and\ \bibinfo {author} {\bibfnamefont {I.~G.}\ \bibnamefont {Savenko}},\
  }\href {\doibase 10.1103/PhysRevB.97.165305} {\bibfield  {journal} {\bibinfo
  {journal} {Phys. Rev. B}\ }\textbf {\bibinfo {volume} {97}},\ \bibinfo
  {pages} {165305} (\bibinfo {year} {2018})}\BibitemShut {NoStop}%
\bibitem [{\citenamefont {Butov}(2003)}]{Butov:2003aa}%
  \BibitemOpen
  \bibfield  {author} {\bibinfo {author} {\bibfnamefont {L.}~\bibnamefont
  {Butov}},\ }\href {\doibase 10.1016/S0038-1098(03)00312-0} {\bibfield
  {journal} {\bibinfo  {journal} {Solid State Comm.}\ }\textbf {\bibinfo
  {volume} {127}},\ \bibinfo {pages} {89 } (\bibinfo {year}
  {2003})}\BibitemShut {NoStop}%
\bibitem [{C2()}]{C2}%
  \BibitemOpen
  \href@noop {} {\ }\BibitemShut {NoStop}%
\bibitem [{\citenamefont {Lerario}\ \emph {et~al.}(2017)\citenamefont
  {Lerario}, \citenamefont {Fieramosca}, \citenamefont {Barachati},
  \citenamefont {Ballarini}, \citenamefont {Daskalakis}, \citenamefont
  {Dominici}, \citenamefont {De~Giorgi}, \citenamefont {Maier}, \citenamefont
  {Gigli}, \citenamefont {K{\'e}na-Cohen},\ and\ \citenamefont
  {Sanvitto}}]{Lerario:2017aa}%
  \BibitemOpen
  \bibfield  {author} {\bibinfo {author} {\bibfnamefont {G.}~\bibnamefont
  {Lerario}}, \bibinfo {author} {\bibfnamefont {A.}~\bibnamefont {Fieramosca}},
  \bibinfo {author} {\bibfnamefont {F.}~\bibnamefont {Barachati}}, \bibinfo
  {author} {\bibfnamefont {D.}~\bibnamefont {Ballarini}}, \bibinfo {author}
  {\bibfnamefont {K.~S.}\ \bibnamefont {Daskalakis}}, \bibinfo {author}
  {\bibfnamefont {L.}~\bibnamefont {Dominici}}, \bibinfo {author}
  {\bibfnamefont {M.}~\bibnamefont {De~Giorgi}}, \bibinfo {author}
  {\bibfnamefont {S.~A.}\ \bibnamefont {Maier}}, \bibinfo {author}
  {\bibfnamefont {G.}~\bibnamefont {Gigli}}, \bibinfo {author} {\bibfnamefont
  {S.}~\bibnamefont {K{\'e}na-Cohen}}, \ and\ \bibinfo {author} {\bibfnamefont
  {D.}~\bibnamefont {Sanvitto}},\ }\href {http://dx.doi.org/10.1038/nphys4147}
  {\bibfield  {journal} {\bibinfo  {journal} {Nature Phys.}\ }\textbf {\bibinfo
  {volume} {13}},\ \bibinfo {pages} {837} (\bibinfo {year} {2017})}\BibitemShut
  {NoStop}%
\end{thebibliography}%
\bibliographystyle{apsrev4-1}

\end{document}